\documentclass[twocolumn]{aastex631}
\usepackage{amsmath}

\let\tablenum\relax %fixes weird error
\usepackage[print-unity-mantissa=false, uncertainty-mode=separate]{siunitx}
\DeclareSIUnit\parsec{pc} % defining parsec unit for use with siunitx
\DeclareSIUnit\year{yr}

\DeclareRobustCommand{\okina}{%
  \raisebox{\dimexpr\fontcharht\font`A-\height}{%
    \scalebox{0.8}{`}%
  }%
}

\received{2025-07-04}
\revised{2025-08-04}
\accepted{2025-08-13}

\newcommand{\oneI}{1I/\okina Oumuamua}
\newcommand{\twoI}{2I/Borisov}
\newcommand{\threeI}{3I/ATLAS}
\newcommand{\OO}{\={O}tautahi--Oxford model}
\newcommand{\OOshort}{\={O}--O model}

\submitjournal{ApJ Letters}

\shorttitle{3I/ATLAS in the \OOshort}
\shortauthors{Hopkins et al.}

\begin{document}

% \title{Oh Lawd He Comin': 3I, the Biggest and Fastest ISO So Far}
% \title{Everybody Duck!!!! 3I, the Biggest and Fastest ISO So Far}
% \title{The Ramans Do Everything In Threes: 3I, the Biggest and Fastest ISO Discovered So Far}
%\title{It's Big(ish) and It's Fast: \threeI{} in the context of the \OO{}}
\title{From a Different Star: \threeI{} in the context of the \={O}tautahi--Oxford interstellar object population model}

\correspondingauthor{Matthew J. Hopkins}
\email{matthew.hopkins@physics.ox.ac.uk}
\author[0000-0001-6314-873X]{Matthew J. Hopkins}
\affiliation{Department of Physics, University of Oxford, Denys Wilkinson Building, Keble Road, Oxford, OX1 3RH, UK}
\affiliation{School of Physical and Chemical Sciences --- Te Kura Mat\={u}, University of Canterbury, Private Bag 4800, Christchurch 8140, New Zealand}

\author[0000-0002-8910-1021]{Rosemary C. Dorsey}
\affiliation{Department of Physics, P.O. Box 64, 00014 University of Helsinki, Finland}

\author[0000-0002-1975-4449]{John C. Forbes}
\affiliation{School of Physical and Chemical Sciences --- Te Kura Mat\={u}, University of Canterbury, Private Bag 4800, Christchurch 8140, New Zealand}

\author[0000-0003-3257-4490]{Michele T. Bannister}
\affiliation{School of Physical and Chemical Sciences --- Te Kura Mat\={u}, University of Canterbury, Private Bag 4800, Christchurch 8140, New Zealand}

\author[0000-0001-5578-359X]{Chris J. Lintott}
\affiliation{Department of Physics, University of Oxford, Denys Wilkinson Building, Keble Road, Oxford, OX1 3RH, UK}

\author[0009-0008-6765-5171]{Brayden Leicester}
\affiliation{School of Physical and Chemical Sciences --- Te Kura Mat\={u}, University of Canterbury, Private Bag 4800, Christchurch 8140, New Zealand}

\begin{abstract}

The discovery of the third interstellar object (ISO), \threeI{} (`3I'), provides a rare chance to directly observe a small body from another Solar System. 
Studying its chemistry and dynamics will add to our understanding of how the processes of planetesimal formation and evolution happen across the Milky Way's disk, and how such objects respond to the Milky Way's potential. 
In this Letter, we present a first assessment of 3I in the context of the \OO{}, which uses data from \emph{Gaia} in conjunction with models of protoplanetary disk chemistry and Galactic dynamics to predict the properties of the ISO population. 
The model shows that both the velocity and radiant of 3I are within the expected range. 
Its velocity predicts an age of over \qty{7.6}{\giga\year} and a high water mass fraction, which may become observable shortly. 
We also conclude that it is very unlikely that 3I shares an origin with either of the previous two interstellar object detections. 

\end{abstract}


\section{Introduction}
\label{sec:intro}
The discovery by the ATLAS survey \citep{Tonry_2018} of the third interstellar object (ISO), \threeI{} (hereafter `3I')\footnote{\url{https://minorplanetcenter.net/mpec/K25/K25N12.html}}, allows us a rare chance to closely study the product of processes of planetesimal formation around another star. 
Though estimates of the population size \citep[$\sim10^{26}$, see e.g.][]{Do2018} suggest that such interstellar objects are the most common macroscopic objects in the Milky Way, their small size and rapid motion has meant that few have yet been observed.

The discovery of \oneI{} and \twoI{} (hereafter `1I' and `2I' respectively) inspired the development of a new field of ISO studies. 
While it had been expected that the ISO population would be primarily cometary in nature \citep{Whipple1975,Francis2005,Jura2011} 1I was apparently inert, had a large axis ratio and tumbling motion \citep{Fraser_2018, Mashchenko2019}, and showed measurable non-gravitational acceleration while leaving the Solar System, possibly caused by outgassing \citep{Micheli_2018,1ITeam2019,Bergner_2023}. 
2I, discovered nearly two years later, was a more conventional, albeit CO- and NH$_2$-rich, comet \citep{Bodewits2020, Deam2025}.
These discoveries inspired work on the origins of ISOs---from scattering in the outer regions of protoplanetary disks \citep{Fitzsimmons2023} to post-main-sequence release of exo-Oort clouds \citep{Levine_2023} to more exotic possibilities, such as fragmentation of exo-Pluto surfaces \citep{Jackson2021} or hydrogen icebergs \citep{Seligman2020}---and their effects on the Milky Way (see, for example, the suggestion by \citet{PfalznerBannister2019} that ISOs incorporated into protoplanetary disks may seed further planet formation). 
A standard picture has emerged, in which planetesimals formed within a protoplanetary disk are scattered by interactions with migrating planets or via stellar flybys, early in the history of a system \citep{Fitzsimmons2023}. 
The number density inferred from observations of the first two ISOs, in addition to studies of scattering in our own Solar System, suggest that such events are common, with $\gtrsim 90\%$ of planetesimals joining the ISO population \citep{Jewitt2023}. 
Such objects spread around the Milky Way's disk in braided streams \citep{Forbes2024}, a small fraction of which intersect our Solar System. 
The observed ISO population is thus truly Galactic, rather than being associated with local stars and stellar populations. 

1I and 2I provided the first observational constraints on models which sought to predict the properties of the ISO population, based on an understanding of their origins, scattering, and subsequent movement within the potential of the Milky Way galaxy \citep{Lintott_2022}. 
This Letter presents the first comparison of the observed properties of 3I with the predictions of the \OO{} \citep{Hopkins2025}, allowing us to understand how this new object is and is not unusual in the context of the putative population from which it is drawn.
As such, it provides an early test of the model, before the 5--50 discoveries of such objects expected from the upcoming Vera C. Rubin Observatory's Legacy Survey of Space and Time (LSST) \citep{Dorsey2025}.

\section{Methods}
\label{sec:methods}

Being products of stars and their planetary systems, the properties of the ISO population of the Milky Way can be predicted from the stellar population.
We do this using the method of \citet{Hopkins2025}, which we outline briefly here.
Using the subset of stars within \qty{200}{\parsec} of the Sun with measured velocity, metallicity and age in \textit{Gaia} DR3 \citep{Gaia2023}, we estimate the solar neighbourhood \textit{sine morte} stellar population.
This is what the stellar population would be if stars did not die, and is needed to predict the ISO population because, unlike stars, ISOs are not removed from the Galactic population at a significant rate.
Using a protoplanetary disk model, we then reweight this \textit{sine morte} stellar distribution in velocity, metallicity and age into a predicted ISO distribution in velocity, composition and age.

The key assumptions of the model, justified in detail in \citet{Hopkins2025}, are that: (i) the number of ISOs released by a star is proportional to its mass and metallicity, (ii) that scattering of ISOs happens efficiently early in each planetary system's lifetime at relatively low speeds (\(\lesssim\qty{10}{\kilo\meter\per\second}\)) and from beyond the water ice line, and (iii) that ISOs are long-lived, outlasting their parent stars which may have undergone stellar death.
Finally, the model assumes (iv) that the current solar neighbourhood \textit{sine morte} stellar population, with appropriate reweighting, can be used as a proxy for the population of ISOs currently passing through the Solar System. This does not assume that these currently-nearby stars are the sources of the ISOs we observe in the Solar System, but instead that the stars and the ISOs trace the same Galactic potential and source populations. Indeed, most observed ISOs will be from distant stars \citep{Forbes2024}

The resulting \={O}tautahi--Oxford (\={O}--O) model predicts correlations between ISO velocity and properties such as composition and age, mostly inherited from similar correlations in the stellar populations. 
These correlations mean that inferences can be made about the physical properties of a discovered ISO from its velocity alone. 
The ISO velocity distribution is highly non-Gaussian and correlated with ISOs' physical properties, and so to predict the properties of a single ISO relative to the model population, only the properties of ISOs from stars with velocities close to the ISO's velocity (within \qty{5}{\kilo\meter\per\second}) are considered when drawing distributions.

3I has a well-determined hyperbolic orbit: the present 67-day long arc\footnote{JPL Horizons heliocentric solution as of 2025-Jul-28: \url{https://ssd.jpl.nasa.gov/tools/sbdb_lookup.html\#/?sstr=C2025N1}} secures its interstellar origin by the key factors of both velocity at infinity \(v_{\infty} = \qty{57.99\pm0.1}{\kilo\meter\per\second}\)
and orbital eccentricity \(e=(6.143\pm0.003)\).
From this orbit fit we infer 3I's Galactic velocity \((U,V,W)= (-51.0, -19.2, 18.5)\,\unit{\kilo\meter\per\second}\) relative to the Sun.
The covariance matrix of the parameters of this orbit fit results in an uncertainty in the velocity of \(<\qty{1.0}{\kilo\meter\per\second}\), mainly in the \(W\)-component, far smaller than the scale of features in the velocity distribution (Fig.~\ref{fig:velocities}). 
This orbit is consistent with the orbit fit to the Vera C. Rubin Observatory observations of 3I \citep{Chandler_2025}.

\section{Chemodynamics and Origin: inferences from the Velocity of \threeI}
\label{sec:results}

\begin{figure*}
    \centering
    \includegraphics[width=0.48\textwidth]{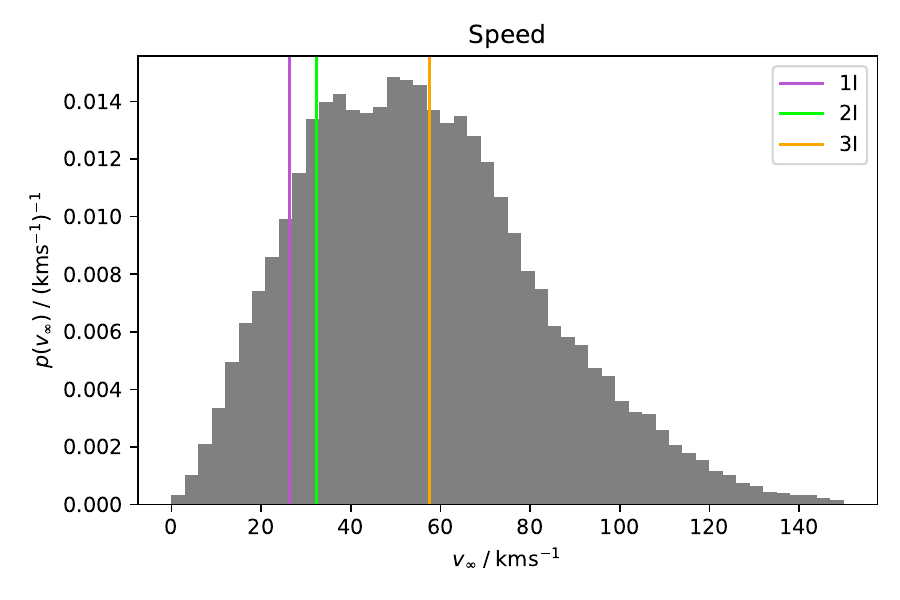}
    \includegraphics[width=0.48\textwidth]{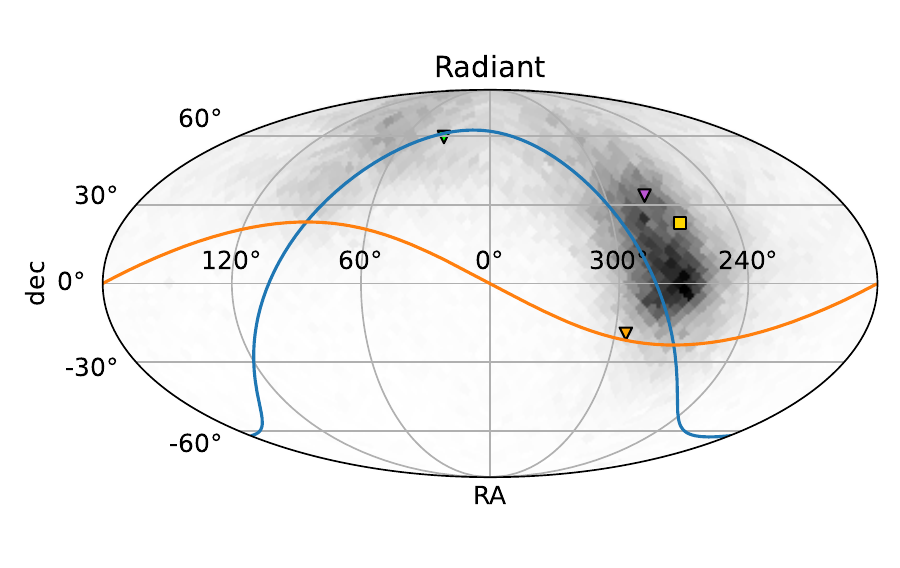}

    \includegraphics[width=0.48\textwidth]{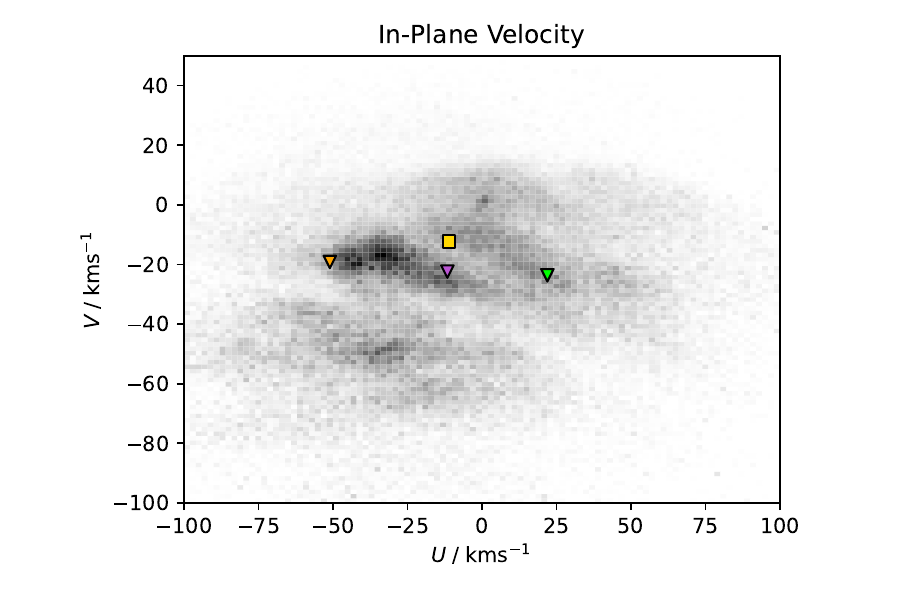}
    \includegraphics[width=0.48\textwidth]{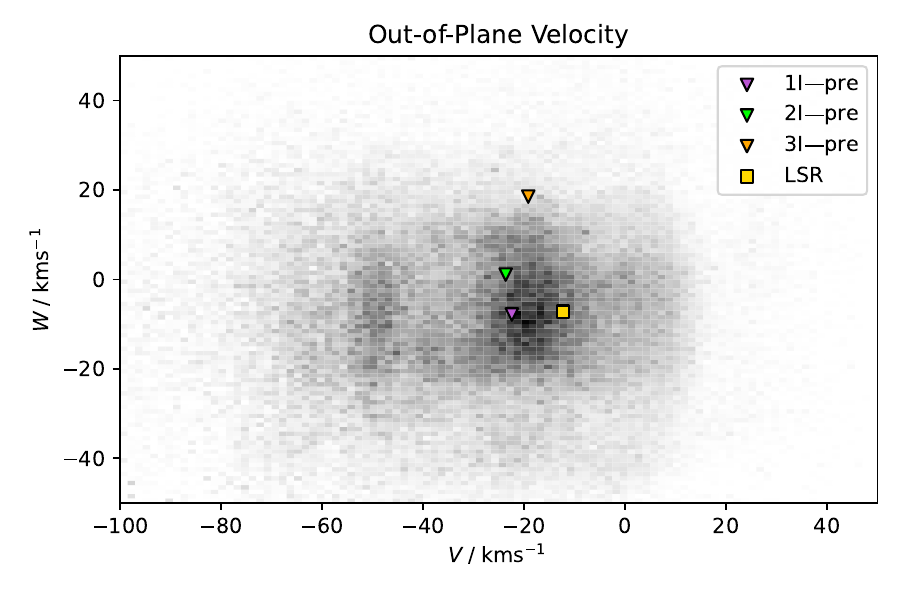}
    \caption{The \OO{}'s predicted asymptotic speed, radiant, and Galactic velocity distribution for \(q<\qty{5}{\astronomicalunit}\) ISOs, overplotted with the three known ISOs. \(U\) points towards the Galactic centre, \(V\) points around the Galactic disk in the direction of the Sun's orbit, and \(W\) points perpendicular to the Galactic plane. The blue and orange lines on the radiant plot mark the Galactic and ecliptic planes respectively, and the yellow square marks the solar apex/local standard of rest \citep{Schonrich_2010}. 
    }
    \label{fig:velocities}
\end{figure*}

Fig.~\ref{fig:velocities} shows the \OOshort's prediction of the distribution of asymptotic speeds \(v_\infty\) and radiants (approach directions) for ISOs on orbits with 
\(q<\qty{5}{\astronomicalunit}\). This range corresponds to the expected perihelia for the upcoming LSST ISO sample \citep{Dorsey2025}, and is also consistent with the perihelia of the three known ISOs.
The distribution of ISOs for different ranges of \(q\) can be seen in \citet{Hopkins2025} Figures B1--3, which show that the only major difference between the distributions is a weighting towards lower speeds for smaller values of \(q\), due to gravitational focussing.
Fig.~\ref{fig:velocities} of this work shows ISOs have a broad distribution of speeds, with which 3I is consistent, though it is faster than its two predecessors.
Radiants are mainly in the Northern hemisphere, though there is a concentration around the equator. 
3I is thus relatively unusual in approaching from so far south (Decl. $\sim -19^\circ$ at discovery).

Fig.~\ref{fig:velocities} also shows the velocity distribution for \(q<\qty{5}{\astronomicalunit}\) ISOs in Cartesian coordinates relative to the Galactic reference frame: \(U\), \(V\), and \(W\).
The velocity distribution is complex and highly non-Gaussian, with large gaps between overdensities in the \(U\)--\(V\) in-plane distribution.
These overdense features, called moving groups, are caused by the dynamical effects of the Galactic spiral arms \citep{Quillen_2011, Hunt_2019, Michtchenko_2018, Ramos_2018, Lucchini_2023}.
These moving groups are not coeval \citep{Nordstrom_2004, Antoja_2008}, and are created continuously as the orbits of both stars and ISOs evolve \citep{Baba2015,Hunt_2018}. 
The pre-encounter velocities of 1I and 2I each lie within a moving group, as is predicted for most ISOs in the \OOshort.
However, 3I's large vertical velocity \(W\) places it outside of the main moving group structures, which are made up of only the low-vertical-motion stars that are more affected by Galactic structure resonances \citep{Daniel2015}.
This large value of \(W\) is the cause of its extreme southern radiant. 
It implies 3I originates from an older and lower-metallicity star, among those on Galactic orbits with large oscillations out of the plane of the Milky Way.

\begin{figure}
    \centering
    \includegraphics[width=\linewidth]{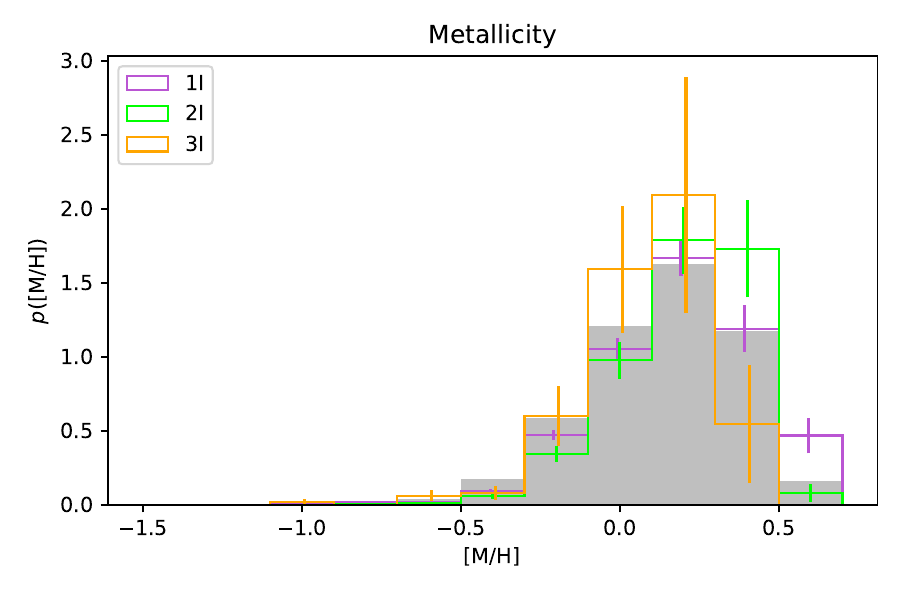}
    \includegraphics[width=\linewidth]{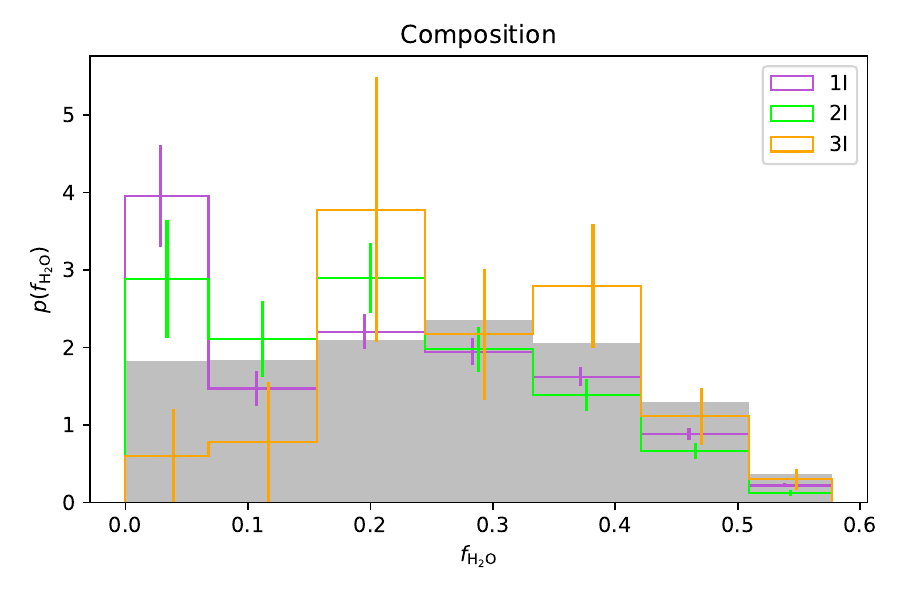}
    
    \includegraphics[width=\linewidth]{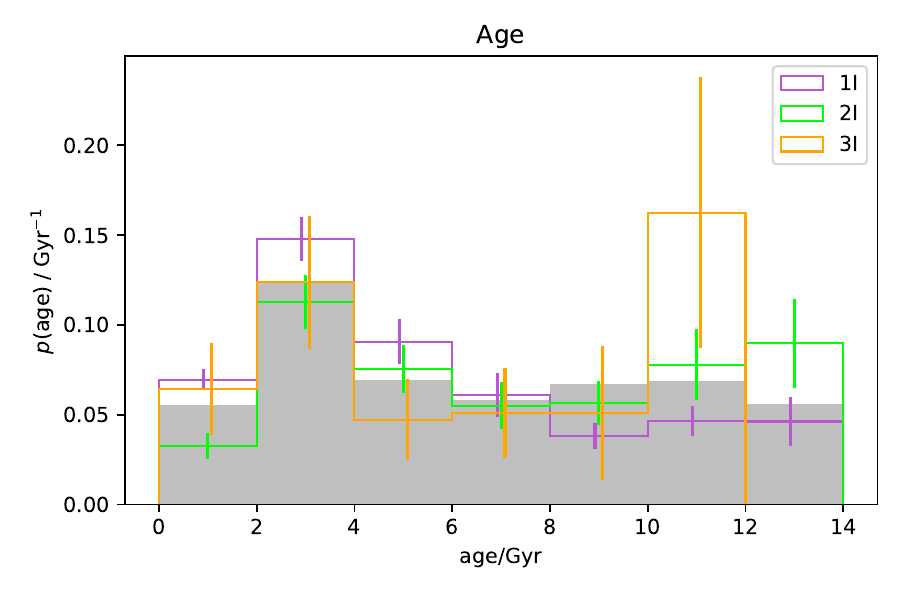}
    \caption{Plots of the predicted parent-star metallicity, water mass fraction and age distributions for the \OO{} \(q<\qty{5}{\astronomicalunit}\) ISOs (grey histograms), overplotted with line histograms of the predicted posterior distributions for ISOs with similar velocity to 1I, 2I and 3I.}
    \label{fig:chemodynamics}
\end{figure}

The top plot of Fig.~\ref{fig:chemodynamics} shows 1I and 2I generally share their velocities with ISOs from higher-metallicity parent stars, but ISOs with 3I's velocity originate around more metal-poor stars. 
From this, we predict in Fig.~\ref{fig:chemodynamics}'s middle plot that ISOs with 3I's velocity are generally more water-rich than the general ISO population.
Such ISOs may exhibit higher water sublimation activity, once close enough to the Sun.
This difference between 3I and the other two known ISOs is due to the difference in moving group membership: stars with velocities within the moving groups have generally higher metallicities than stars with velocities outside.

Although the \OOshort{} predicts some weak correlations between age and velocity, ISOs of all ages can be found with the velocities of each of 1I, 2I and 3I. 
Looking at the predicted age distributions of ISOs at each known ISO's velocity, Fig.~\ref{fig:chemodynamics} (lower plot) shows that 1I shares its velocity with slightly younger (i.e. \(<\qty{6}{\giga\year}\)) ISOs than the overall ISO distribution, and similarly 2I shares its velocity with slightly older (i.e. \(>\qty{10}{\giga\year}\)) ISOs.
Despite its large out-of-plane (vertical) velocity shown in Fig.~\ref{fig:velocities}, at the velocity of 3I we predict ISOs to have a wide spread of ages, with no clear or strong skew towards either younger or older ages.
The lack of a strong correlation between velocity and age means that any age estimate for an ISO based only on its velocity must have a large uncertainty.
It is important that future works on 3I, and on ISOs yet to be discovered, take this into account.\footnote{In an early version of this manuscript, distributed on arXiv, we explicitly identified the origin of 3I with the thick disk of the Milky Way. Though there is no clear transition between thin and thick disk, and the nature of the latter remains a topic of some dispute \citep[e.g.][and references therein]{Bovy_2012,Duong_2018}, to avoid doubt the present text reflects the definition of the thick disk used by \citep{Gilmore1983}, which applies only to objects further than 1kpc from the Galactic plane, and does not include 3I. We thank Eric Mamajek for pointing out the lack of clarity in our original text.}

\section{Age estimates for \threeI{} using the age--velocity dispersion relation}
\label{sec:AVR}
Though the correlation between the velocity of a single object and its age is weak, there is a well-established result that the velocity dispersion of populations of objects in the Milky Way disk increases with age. 
This so-called age--velocity dispersion relation (AVR) allows the age of an object to be estimated. 
The AVR is thought to be caused by repeated scatterings over time by large-scale structures, such as the spiral arms and giant molecular clouds \citep{Gustafsson2016}.
It should be noted that the contribution from scattering of close passages with other stars is negligible, due to the long scattering timescale of \(\qty{1e7}{\giga\year}\) \citep{Forbes2024}, so the large velocity of 3I is not a indication that it has undergone a close encounter with a star before the Sun.

\begin{figure}
    \centering
    \includegraphics[width=\linewidth]{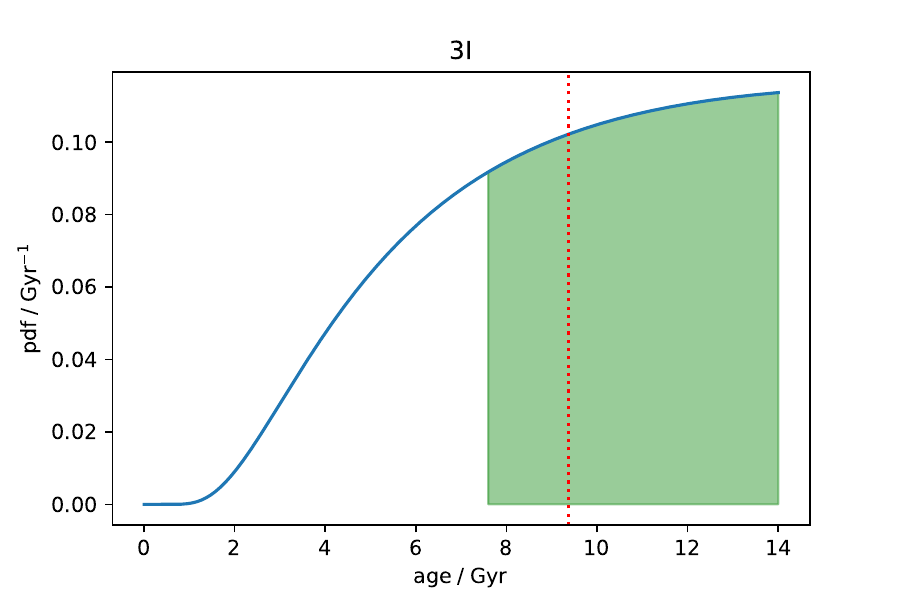}
    \caption{Age posterior distribution for 3I, based on its out-of-plane velocity \(W\), with shaded 68\% confidence intervals. 
    Vertical line marks biased point estimates, which generally underestimate the age, especially in the case of 1I (see Fig.~\ref{fig:ageDists1I2I}, Appendix).}
    \label{fig:ageDists}
\end{figure}

Here, we use a simplified version of the method of \citet{Almeida-Fernandes2018} to provide confidence intervals on the age of an ISO, showing the calculated age posterior distribution for 3I in Fig.~\ref{fig:ageDists}, and 1I and 2I in Fig.~\ref{fig:ageDists1I2I}. 
\citet{Almeida-Fernandes2018}'s method models the 3D velocity distribution of ISOs as a Gaussian with dispersion increasing with age.
However, as shown in the \(U\)--\(V\) plot of Fig.~\ref{fig:velocities}, the actual stellar or ISO velocity distribution is not well approximated by a Gaussian distribution\footnote{Further discussed in \citet{Hopkins2025}.}: there are large gaps and over-densities, and the LSR neither lies near the centre of the velocity distribution nor at a peak. 
In order to avoid the highly non-Gaussian \(U\)--\(V\) velocity distribution, we only use the out-of-plane component \(W\). 
The \(W\) distribution is also not perfectly Gaussian; however it is at least centred on the \(W\) component of the LSR (\(W_\mathrm{LSR}\)) and does not have large overdensities and gaps like the \(U\)--\(V\) distribution.
Therefore we assume that the ISO \(W\) distribution is given by a 1D Gaussian centred on \(W_\mathrm{LSR}\) with dispersion \(\sigma_W\) increasing with age as described by the two-slope AVR as \citet{Almeida-Fernandes2018}, described in their Table 3 and Eq.~8.
With this we calculate a likelihood distribution for each ISO's out-of-plane-velocity \(W\) at each age in the range 0.01--14~\unit{\giga\year}, which under the assumption of a uniform age prior is equal to the posterior distribution for each ISO's age, given its out-of-plane velocity. 

The resulting posterior distributions for each known ISO's age is plotted along with a 68\% confidence interval in Figs.~\ref{fig:ageDists} and \ref{fig:ageDists1I2I}.
These intervals are 0.01--7.2 \unit{\giga\year} for 1I, 0.36--8.8 \unit{\giga\year} for 2I, and 7.6--14 \unit{\giga\year} for 3I.
The large width of these intervals reflect the large uncertainties present in these age estimates when the AVR is treated properly.
Corroborating our predictions in \S~\ref{sec:results}, each posterior remains non-zero for almost all values of age --- meaning based on \(W\), each ISO could potentially be of any age.
However, in this approach 3I is generally older than either 1I or 2I.
The only strong constraint which can be placed is where the posterior for 3I approaches zero for ages \(<\qty{1}{Gyr}\).\footnote{After submission of this manuscript, \citet{Taylor_2025} was listed as an arXiv pre-print; they use a different age--velocity dispersion relation to estimate the age of each ISO, finding age ranges consistent with ours.} 

One method previously used to provide point estimates of the ages of 1I and 2I is to equate the velocity of the ISO relative to the LSR with a velocity dispersion, before reading off the corresponding age from the AVR \citep[e.g.][]{Feng2018,Hallatt2020,Hsieh2021}, but it is clear from Figs.~\ref{fig:ageDists} and \ref{fig:ageDists1I2I} that using this estimate of the velocity distribution will result in a biased, typically underestimated age\footnote{This is especially true for 1I, for which applying such a process produces a very low age estimate (Fig.~\ref{fig:ageDists1I2I}).}. 
This is important because in order for an ISO to be traceable back to its parent star, it must have an extremely young age of \(<\qty{10}{\mega\year}\) \citep{Zhang2018,Zuluaga2018}.
Several works \citep{Dybczynski_2018,Dybczynski_2019,Bailer-Jones_2018,Bailer-Jones_2020} attempted this back-tracing for both 1I and 2I and listed `feasible' parent star candidates, but such a procedure relies on an underestimated age. 
Tracing the origin of 3I back to a single star or cluster is equally unfeasible.

\section{Probability of \threeI{} having a common origin with 1I or 2I}
\label{sec:twins}

\begin{figure}
%    \centering
    \includegraphics[width=\linewidth]{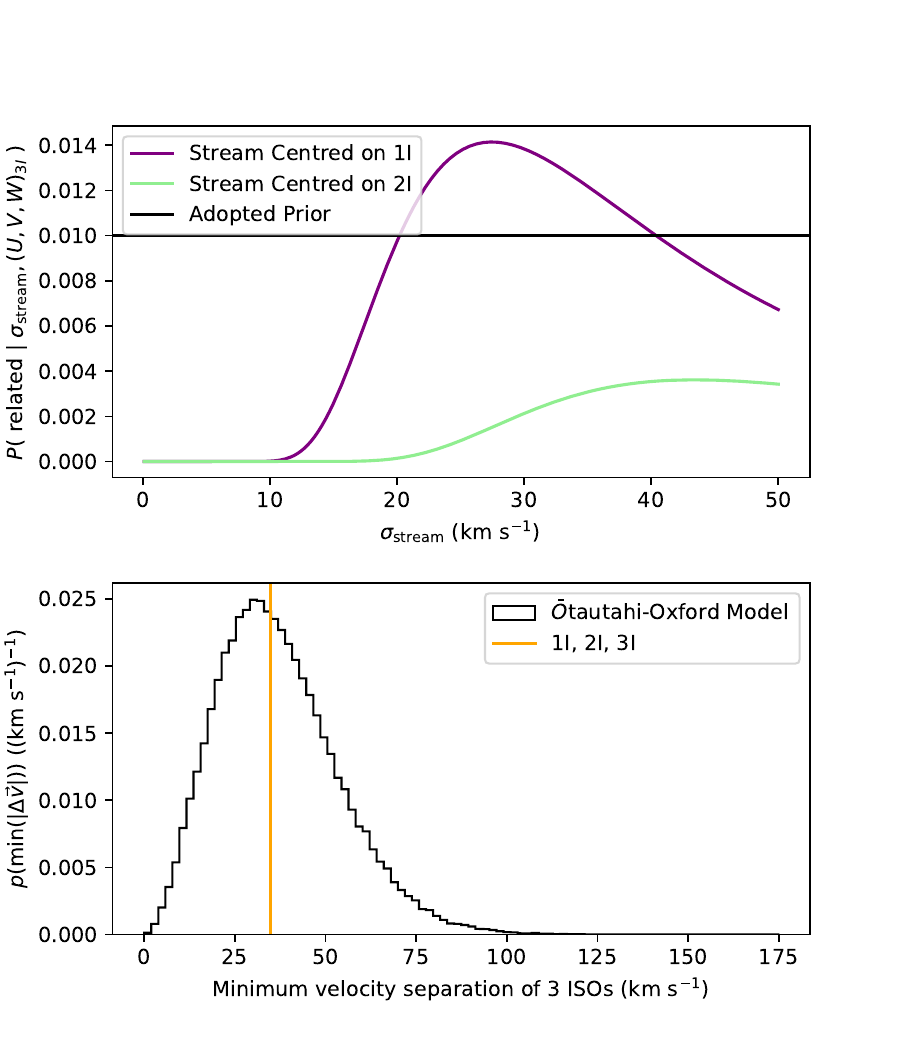}
    \caption{The improbability of an association between 3I and a previous ISO. The top panel shows the Bayesian posterior probability that 3I is related to 1I (purple line) or 2I (light green line), conditioned on a known velocity dispersion, $\sigma_\mathrm{s}$, of the ISO stream centred on 1I or 2I respectively. The probability is low, less than 1.4\%, regardless of $\sigma_\mathrm{s}$. The bottom panel shows the distribution of the minimum velocity separation for random triplets of ISOs drawn from the \OO{}, assuming no common origins for any ISOs. The actual separation in velocity of the observed ISOs (orange line) is in the central part of the distribution, indicating that there is minimal evidence for an association.}
    \label{fig:twins}
\end{figure}

Because ISOs orbit the Galaxy in streams \citep{PortegiesZwart2021}, it is likely that we will eventually see ISOs with a common origin, either from the same star or from the same star cluster \citep{Forbes2024}. 
We therefore examine the possibility that 3I shares a common parent system with either 1I or 2I. 

First, we apply Bayes' Theorem to estimate the probability that 3I is associated with one of the two other known ISOs. 
We assume that the typical spread in velocities within the ISO stream associated with either 1I or 2I is known, $\sigma_\mathrm{s}$. 
Following an approach similar to \citet{Forbes2019}, we have:
\begin{equation}
\begin{split}
    &P(\mathrm{r}|\sigma_\mathrm{s}, \vec{v}_{3I}) \\ &= \frac{p(\vec{v}_{3I}|\mathrm{r},\sigma_\mathrm{s})P(\mathrm{r}|\sigma_\mathrm{s})}{p(\vec{v}_{3I}|\mathrm{r},\sigma_\mathrm{s})P(\mathrm{r}|\sigma_\mathrm{s}) + p(\vec{v}_{3I}|\mathrm{nr},\sigma_\mathrm{s})(1-P(\mathrm{r}|\sigma_\mathrm{s}))}
\end{split}
\end{equation}
where we have abbreviated `related' to `r', `unrelated' to `nr' and `stream' to `s'.
We estimate each of the factors as follows. 

We adopt the prior $P(\mathrm{r}|\sigma_\mathrm{s}) = 0.01$, which is close to the highest probability of having a related ISO among the models considered in \citet{Forbes2024}. In particular, models with this probability or higher are all associated with ISOs that come from the same {\em cluster} -- for ISOs that come from the same star, the prior probability is considerably lower. The prior may also be substantially modulated by the given value of $\sigma_\mathrm{s}$ for ISOs that come from the same {\em star}, because if the ISO comes from the same star, it is more likely to have a lower $\sigma_\mathrm{s}$. However, we neglect this complication here.

For the likelihood, $p(\vec{v}_{3I}|\mathrm{r},\sigma_\mathrm{s})$, we assume that the putative stream associated with either 1I or 2I, is centred on the velocity of 1I or 2I respectively. This is the simplest assumption to make, and avoids the additional complication of integrating over many possible stream central velocities. The likelihood is therefore a tri-variate Gaussian centred on the velocity of 1I or 2I (we consider them separately), with an isotropic covariance matrix equal to $\sigma_\mathrm{s}^2$ times the 3$\times$3 identity matrix. This particularly simple form could be modified to respect the contours of the moving groups, for instance, but again we avoid this complication.

Finally, to estimate the probability density of $\vec{v}_{3I}$ under the assumption that 3I is unrelated to the other two ISOs, we compute the KDE based on the \(U\),\(V\),\(W\) Galaxy-oriented Cartesian velocities of $10^6$ sample draws from the \OOshort{} (\S~\ref{sec:methods}). 
We adopt the \citet{Silverman1986} bandwidth for the KDE, and then evaluate the KDE at the velocity of 3I. The results are shown in the top panel of Figure \ref{fig:twins}. The posterior probability is low regardless of $\sigma_\mathrm{s}$, barely exceeding the prior in the case of an association with 1I, and never exceeding the prior for an association with 2I.

This result is easily understood by examining the distribution of velocity separations when sets of 3 ISOs are drawn from the Gaia-based model of \citet{Hopkins2025}. For each triplet of ISOs, we compute the minimum magnitude of the velocity difference among the 3 pairs of ISOs in the triplet. That is, we compute the minimum of $|\vec{v}_{1}-\vec{v}_2|$, $|\vec{v}_{1}-\vec{v}_3|$, and $|\vec{v}_{2}-\vec{v}_3|$, where the subscript indicates which of the 3 ISOs in the draw is being considered. The distribution of this test statistic is shown in the bottom panel of Figure \ref{fig:twins}, along with a vertical line indicating the observed value for the velocities of 1I, 2I, and 3I. The observed value is in the middle of the distribution, indicating that no additional clumping beyond what is already present in the \citet{Hopkins2025} model is necessary to explain the separation in velocities, and so there is no kinematic evidence that 3I comes from the same star or star cluster as either of the previous ISOs. 
This is to be expected given the intrinsically low probability of seeing an ISO from the same stream, given the large number of streams through which the Sun is likely currently passing \citep{Forbes2024}.

\section{Unusual Aspects of \threeI{}}
\label{sec:sizeDist}

Early reports show that at its current heliocentric distance $r_h\sim 4.4$~au, 3I is a quietly active comet with a compact coma \citep{seligman2025, ATel_17263, ATel_17264}; hence the absolute magnitude of 3I from the earliest photometry ($H_V=11.9\pm0.6$) is an upper limit on the true nucleus size.\footnote{After submission of this manuscript, \citet{Chandler_2025} was listed as an arXiv pre-print; they estimate a nucleus absolute magnitude of \(H_V=13.7\pm0.2\), corresponding to a radius of \(5.6\pm0.7\,\unit{\kilo\meter}\).} 
The larger inferred size of 3I (on the order of $\lesssim\qty{10}{\kilo\meter}$) compared to 1I and 2I ($\sim0.1$--1~\unit{\kilo\meter}) may constrain the size-frequency distribution (SFD) of the intrinsic galactic ISO population in interesting ways.
Estimation of the intrinsic SFD slope based on the detections of 1I and 2I suggests a steep value \citep[$q\sim3$-4,][]{Jewitt2020}, i.e. small ISOs are much more common than large ISOs in the Galaxy.
Additionally, simulations of future ISO discoveries within LSST show that their expected median size is strongly dependent on the intrinsic SFD slope, with steeper slopes resulting in fewer and smaller discovered objects \citep{Marceta2023, Dorsey2025}.
This also holds for surveys shallower than LSST (limiting magnitude $m_r\sim24$), as their limiting magnitude impacts the smallest possible size of discoverable ISO, rather than their distribution.
Hence, the discovery of a larger ISO suggests that the integrated Galactic ISO SFD may in fact be shallower than previously expected.

This also aligns with tentative number density estimates for objects larger than 3I, $n\left( H \leq 12\right)\sim 10^{-3}$~au$^{-3}$ \citep{seligman2025}.
Their value ignores several survey selection effects and survivorship bias, and should therefore be treated cautiously as an order-of-magnitude estimate until full survey simulation of the detection of 3I in ATLAS can be carried out (Dorsey et al., in prep); however, it can be used to roughly estimate the slope of the ISO SFD.
Using the number density estimates for objects larger than 1I \citep[$n\left(H\leq22\right)\sim10^{-1}$~au$^{-3}$;][]{Do2018} and 3I, the ISO SFD can be modelled empirically by a power law in \(H\) with slope of \(\alpha\sim0.2\), equivalent at fixed albedo to a slope in diameter of $q\sim2$.
While \citet{Dorsey2025} demonstrated that a single-slope power law with $q<2.5$ is not physically feasible for ISOs (indicating that a multi-slope power law is necessary instead), this extrapolation does suggest that the SFD of ISOs is likely shallower than was previously predicted using 1I and 2I, or perhaps wavy like other Solar System populations \citep{Bottke2023}.

The orbital orientation of 3I is also unanticipated.
Due to the Galactic motion of the Sun, ISOs preferentially infall to the Solar System from the direction of the Solar Apex in the North Celestial Hemisphere \citep{McGlynn1989, Seligman2018}.
Observatories based in the Southern Hemisphere are therefore observationally biased to discover objects on orbits with perihelia in the Southern Celestial Hemisphere, i.e. arguments of perihelia $180^\circ\leq\omega\leq360^\circ$.
The ATLAS Chile observatory and the Rubin Observatory are located at similar latitudes ($30.47^\circ$ and $30.24^\circ$~S respectively) and can be assumed to have rather similar pointing biases.
\citet{Dorsey2025} used the \OOshort{} to determine that while all orbital $\omega$ of ISOs are discoverable with non-zero probability in LSST, orbits with perihelia in the Southern Hemisphere dominate the discovered population ($\sim70\%$ of objects).
Of the orbits with perihelia in the Northern Hemisphere ($0^\circ\leq\omega\leq180^\circ$), those with argument of perihelia $60^\circ\leq\omega\leq150^\circ$ were most strongly biased against discovery.
Specifically, the likelihood that an ISO discovered in LSST will have an orbit similar to 3I ($125^\circ \leq \omega \leq 130^\circ$) is $\sim1\%$.
By comparison, the arguments of perihelia\footnote{$241.8105^\circ\pm0.0012^\circ$ and $209.12369^\circ\pm0.00011^\circ$ respectively; \url{https://ssd.jpl.nasa.gov/tools/sbdb_lookup.html\#/?sstr=1i} and \url{https://ssd.jpl.nasa.gov/tools/sbdb_lookup.html\#/?sstr=2i}} for 1I and 2I are a factor of $\sim$3--4 times more discoverable in the same survey. 
The discovery of an ISO on an orbit like that of 3I by a Southern Hemisphere telescope could simply be a rare happenstance (an unlikely but not impossible random draw from the \OOshort), a yet-unmodelled ISO stream encountering the Solar System, or perhaps evidence toward a different model of Galactic ISO orbits.

\section{Conclusion}

The third known ISO, \threeI, has physical differences to the first two ISOs detected, but its orbit and incoming velocity place it as a member of the Galactic population of ISOs.
We find that though \threeI{} is entering the Solar System at a higher speed than 1I or 2I, this is still well within the expected distribution of ISO speeds inferred from the \OO{}.
The model predicts that ISOs with 3I's velocity are generally more water-rich.
While at 3I's present $r_h$ water would be challenging to detect, this should be observable in its coma once 3I crosses into water-ice sublimation distances within $\sim 3$~au.

\threeI's direction of approach is more southerly than the bulk of expected ISOs the Sun will encounter, due to its very large vertical motion out of the Galactic plane. 
From its kinematics, we rule out the possibility that \threeI{} comes from the same star, or same cluster, as either 1I or 2I, but the velocity does tell us about its origin: it shares its Galactic orbit with older and lower-metallicity stars. Though the constraints are not tight and an exact determination of age is not possible, our models suggest 3I's age is likely to be over \qty{7.6}{\giga\year}, older than 1I, 2I, and all Solar System objects.

The rich evidence that an ISO's velocity contains in its chemodynamics demonstrate the benefit of rapid orbital arc determination by the community. 
ISOs provide the opportunity to gain evidence of the process of planetesimal formation and evolution from a host of Galactic environments, and further observations of \threeI{} will allow us to constrain and test the assumptions made in the \OO{}.

\appendix
\section{Age posteriors for 1I and 2I}
\begin{figure}
    \centering
    \includegraphics[width=0.45\linewidth]{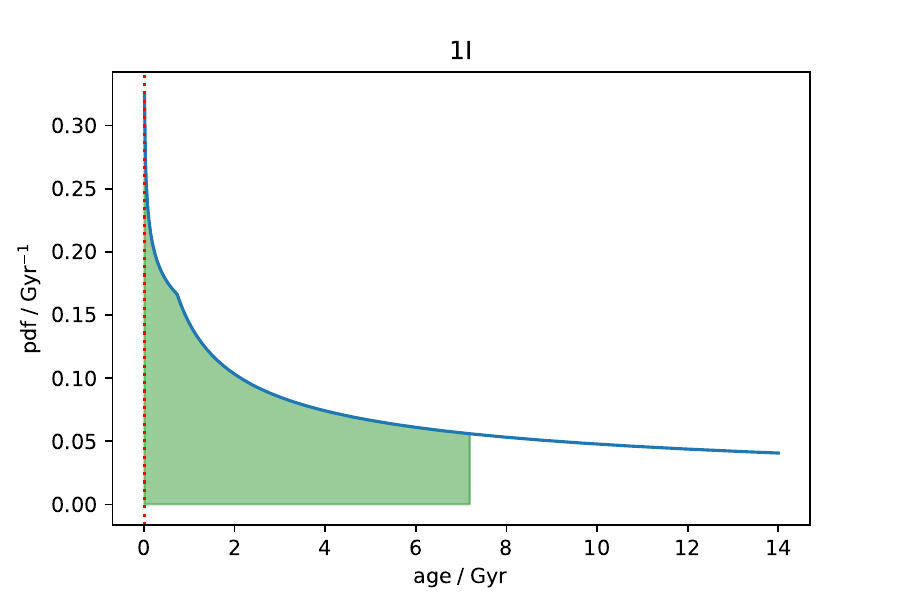}
    \includegraphics[width=0.45\linewidth]{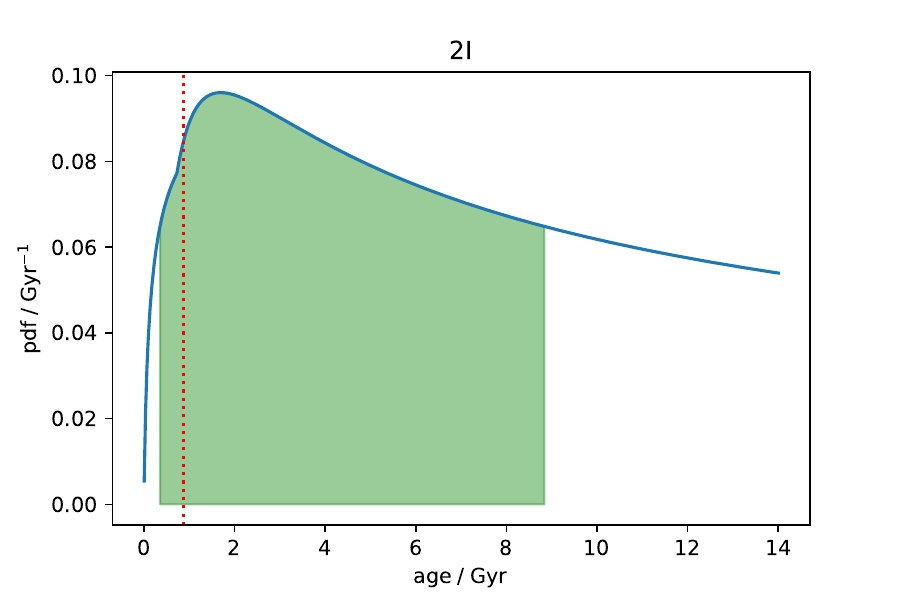}
    \caption{Age posterior distribution for 1I and 2I, based on their out-of-plane velocity \(W\), with shaded 68\% confidence intervals. 
    Vertical line marks biased point estimates, which generally underestimate the age, especially in the case of 1I.}
    \label{fig:ageDists1I2I}
\end{figure}
Fig.~\ref{fig:ageDists1I2I} shows the posterior distributions for 1I and 2I, analogous to that of 3I in Fig.~\ref{fig:ageDists}. 
With 68\% confidence intervals of 0.01–7.2 Gyr and 0.36–8.8 Gy respectively, they demonstrate that ISO age estimates from velocity are high in uncertainty, and point estimates made by equating the velocity to a velocity dispersion in the AVR generally give biased results, underestimating an ISO's age.

\begin{acknowledgements}
M.J.H. acknowledges support from the Science and Technology Facilities Council through grant ST/W507726/1, and the Oxford Physics Cricket Club.
R.C.D. acknowledges support from grant \#361233 awarded by the Research Council of Finland to M. Granvik.
M.T.B. and J.C.F. appreciate support by the Rutherford Discovery Fellowships from New Zealand Government funding, administered by the Royal Society Te Ap\={a}rangi.
M.T.B. appreciates discussions on water detectability with Cyrielle Opitom, Matthew Knight, Manuela Lippi, Brian Murphy and Emmanuel Jehin.
C.J.L. thanks Alice Stothart and Sebastian Thornton for assistance with drafting this paper.

\end{acknowledgements}

\bibliography{export-bibtex}
\bibliographystyle{aasjournal}

\end{document}